\documentclass[pra,amsmath,amssymb,amsfonts,twocolumn]{revtex4}
\usepackage{graphicx}%
\usepackage{dcolumn}
\usepackage{amsmath}
\usepackage[latin1]{inputenc}

\begin{document}

\title{Silica-on-Silicon Waveguide Quantum Circuits}
\author{Alberto Politi, Martin J. Cryan, John G. Rarity, Siyuan Yu, and Jeremy L. O'Brien}
\affiliation{Centre for Quantum Photonics, H. H. Wills Physics Laboratory \& Department of Electrical and Electronic Engineering, University of Bristol, Merchant Venturers Building, Woodland Road, Bristol, BS8 1UB, UK}


\begin{abstract}
Quantum technologies based on photons are anticipated in the areas of information processing, communication, metrology, and lithography. While there have been impressive proof-of-principle demonstrations in all of these areas, future technologies will likely require an integrated optics architecture for improved performance, miniaturization and scalability. We demonstrated high-fidelity silica-on-silicon integrated optical realizations of key quantum photonic circuits, including two-photon quantum interference with a visibility of $94.8\pm0.5\%$; a controlled-NOT gate with logical basis fidelity of $94.3\pm0.2\%$; and a path entangled state of two photons with fidelity $>92\%$. 
\end{abstract}

\maketitle


Quantum information science \cite{nielsen} has shown that harnessing quantum mechanical effects can dramatically improve performance for certain tasks in communication, computation and measurement. However, realizing such quantum technologies is an immense challenge, owing to the difficulty in controlling quantum systems and their inherent fragility. Of the various physical systems being pursued, single particles of light---photons---are often the logical choice, and have been widely used in quantum communication \cite{gi-rmp-74-145}, quantum metrology \cite{na-sci-316-726,hi-nat,ob-sci-318-1393}, and quantum lithography \cite{ka-oe-15-14249} settings. Low noise (or decoherence) also makes photons attractive quantum bits (or qubits), and they have emerged as a leading approach to quantum information processing \cite{ob-sci-318-1567}.

In addition to single photon sources \cite{sps} and detectors \cite{spd}, photonic quantum technologies will rely on sophisticated optical circuits involving high-visibility classical and quantum interference. Already a number of photonic quantum circuits have been realized for quantum metrology \cite{na-sci-316-726,hi-nat,mi-nat-429-161,wa-nat-429-158,re-prl-98-223601,su-pra-74-033812}, lithography \cite{ka-oe-15-14249}, quantum logic gates \cite{pi-pra-68-032316,ob-nat-426-264,ob-prl-93-080502,la-prl-95-210504,ki-prl-95-210505,ok-prl-95-210506,ga-prl-93-020504}, and other entangling circuits \cite{wa-nat-434-169,ki-prl-95-210502,pr-nat-445-65,lu-nphys-3-91}. However, these demonstrations have relied on large-scale (bulk) optical elements bolted to large optical tables, thereby making them inherently unscalable and confining them to the research laboratory. In addition, many have required the design of sophisticated interferometers to achieve the sub-wavelength stability required for reliable operation.

We demonstrated the fundamental building blocks of photonic quantum circuits using silica waveguides on a silicon chip: high visibility ($98.5\pm 0.4\%$) classical interference; high visibility ($94.8\pm0.5\%$) two photon quantum interference; high fidelity controlled-NOT (CNOT) entangling logic gates (logical basis fidelity $F=94.3\pm0.2\%$); and on-chip quantum coherence confirmed by high fidelity ($>92\%$) generation of a  two-photon path entangled state. The monolithic nature of these devices means that the correct phase can be stably realized in what would otherwise be an unstable interferometer, greatly simplifying the task of implementing sophisticated photonic quantum circuits. We fabricated 100's of devices on a single wafer and find that performance across the devices is robust, repeatable and well understood. 

A typical photonic quantum circuit takes several optical paths or ``modes" (some with photons, some without) and mixes them together in a linear optical network, which in general consists of `nested' classical and quantum interferometers (\emph{eg.} Fig 1C). In a standard bulk optical implementation the photons propagate in air, and the circuit is constructed from mirrors and beamsplitters (BSs), or ``half reflective mirrors", which split and recombine optical modes, giving rise to both classical and quantum interference. High visibility quantum interference \cite{ho-prl-59-2044} 
demands excellent optical mode overlap at a BS, which requires exact alignment of the modes; while high visibility classical interference also requires sub-wavelength stability of optical path lengths, which often necessitates the design and implementation of sophisticated stable interferometers. Combined with photon loss, interference visibility is the major contributor to optical quantum circuit performance. 

In conventional (or classical) integrated optics devices light is guided in waveguides---consisting of a ``core" and slightly lower refractive index ``cladding" (analogous to an optical fiber)---which are usually fabricated on a semiconductor chip. By careful choice of core and cladding dimensions and refractive index difference it is possible to design such waveguides to support only a single transverse mode for a given wavelength range. Coupling between waveguides, to realize BS-like operation, can be achieved when two waveguides are brought sufficiently close together that the evanescent fields overlap; this is known as a directional coupler. By lithographically tuning the separation between the waveguides and the length of the coupler the amount of light coupling from one waveguide into the other (the coupling ratio $1-\eta$, where $\eta$ is equivalent to BS reflectivity) can be tuned. 

The most promising approach to photonic quantum circuits for practical technologies appears to be realizing integrated optics devices which operate at the single photon level. However, there has been no demonstration of quantum behavior of light in waveguide devices. Key requirements are: single mode guiding of single photons; high visibility classical interference; high visibility quantum interference; and the ability to combine these effects in a waveguide optical network. 

To achieve these goals we required a material system that is (1) low loss at a wavelength of $\lambda\sim800$ nm, where commercial silicon avalanche photodiode single photon counting modules (SPCMs) are near their peak efficiencey of $\sim$70\%, (2) enables a refractive index contrast $\Delta=(n_{core}^2-n_{cladding}^2)/2n_{core}^2$ that results in single mode operation for waveguide dimensions comparable to the core size of conventional single mode optical fibers at $\sim$800 nm (4-5 $\mu$m), to allow good coupling of photons to fiber-coupled single photon sources and detectors, and (3) is amenable to standard optical lithography fabrication techniques. The most promising material system to meet these requirements was silica (silicon dioxide SiO$_2$), with a low level of doping to control the refractive index, grown on a silicon substrate (Fig. 1B).

\begin{figure}[t]
\vspace{-1cm}
\begin{center}
\includegraphics*[width=0.48\textwidth]{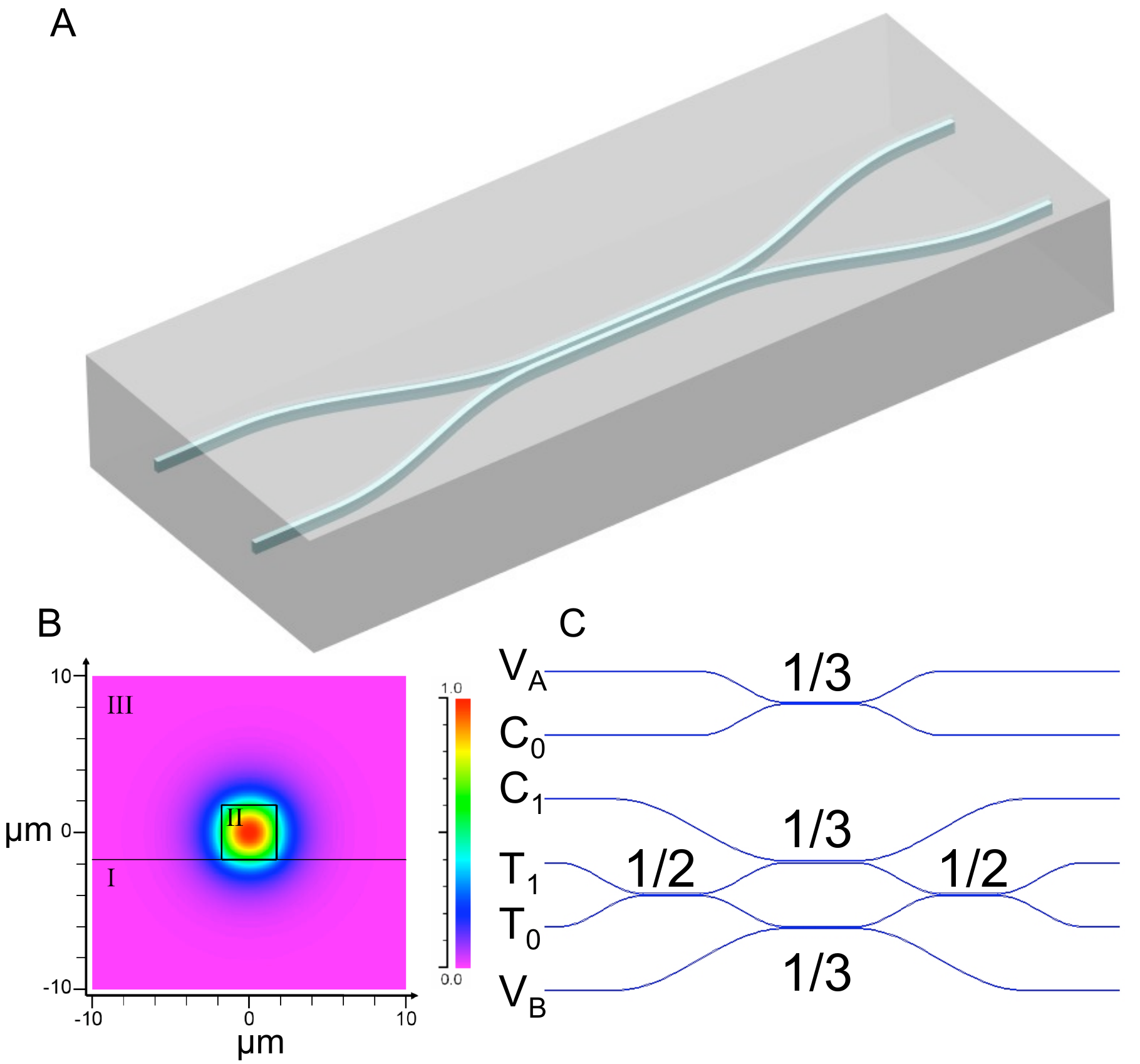}
\vspace{-0.2cm}
\caption{Silica-on-silicon integrated quantum photonic circuits.  ({\bf A}) A direction coupler, which can be used as the building block for integrated photonic quantum circuits by replacing the bulk BS. ({\bf B}) The modeled transverse intensity profile of the guided mode superimposed on the waveguide structure. ({\bf C}) Design of the integrated two-photon controlled-NOT quantum logic gate.}
\label{fig1}
\end{center}
\vspace{-0.8cm}
\end{figure}

Having chosen silica as a material system, we selected a refractive index contrast of $\Delta=0.5$\% to give single mode operation at 804 nm for $3.5\times 3.5$ $\mu$m waveguides (as determined by modeling with the vectorial mode solving package Fimmwave). This value of $\Delta$ provides moderate mode confinement (the transverse intensity profile is shown in Fig. 1B)  thereby minimizing the effects of fabrication or modelling imperfections. We designed a number of devices including directional couplers with various $\eta$'s, Mach-Zender interferometers (consisting of two directional couplers), and more sophisticated devices built up from several directional couplers with different $\eta$'s using Rsoft's beam propagation method (BPM) package. 

Starting with a 4" silicon wafer, a 16 $\mu$m layer of thermally grown undoped silica was deposited as a buffer (material I in Fig. 1B), followed by flame hydrolysis deposition of a 3.5 $\mu$m waveguide core of silica doped with germanium and boron oxides (II). The core material was patterned into 3.5 $\mu$m wide waveguides via standard optical lithography techniques, 
and finally overgrown with a further 16 $\mu$m cladding layer of  phosphorus and boron doped silica with a refractive index matched to that of the buffer (III). The wafer was diced into several dozen individual chips, each containing typically several devices. Some chips were polished to enhance coupling in and out of the waveguides \footnote{All devices were fabricated at the Centre for Integrated Photonics}.

We used a BBO type-I spontaneous parametric downconversion (SPDC) crystal, pumped with a 60 mW 402 nm continuous wave diode laser to produce 804 nm degenerate photon pairs at a detected rate of ~4000 s$^{-1}$ when collected into single mode polarization maintaining fibers (PMFs). We used 2 nm interference filters to ensure good spectral indistinguishability \footnote{In a separate experiment with a bulk optics BS we used this source to observe quantum interference with $V$=97\%}. Single photons were launched into the waveguides on the integrated optical chips and then collected at the outputs using two arrays of 8 PMFs, with 250 $\mu$m spacing, to match that of the waveguides, and detected with fiber coupled SPCMs. The PMF arrays and chip were  directly buttcoupled, with index matching fluid. Overall coupling efficiencies of $\sim$60\% through the device (insertion loss=40\%) were routinely achieved \footnote{We note that minimal effort was made to match the waveguide and fiber modes---no tapers were used for example---and that the coupling efficiency was likely limited by a mismatch of mode size and shape.}.

\begin{figure}[t]
\begin{center}
\vspace{-1cm}
\includegraphics[width=0.4\textwidth]{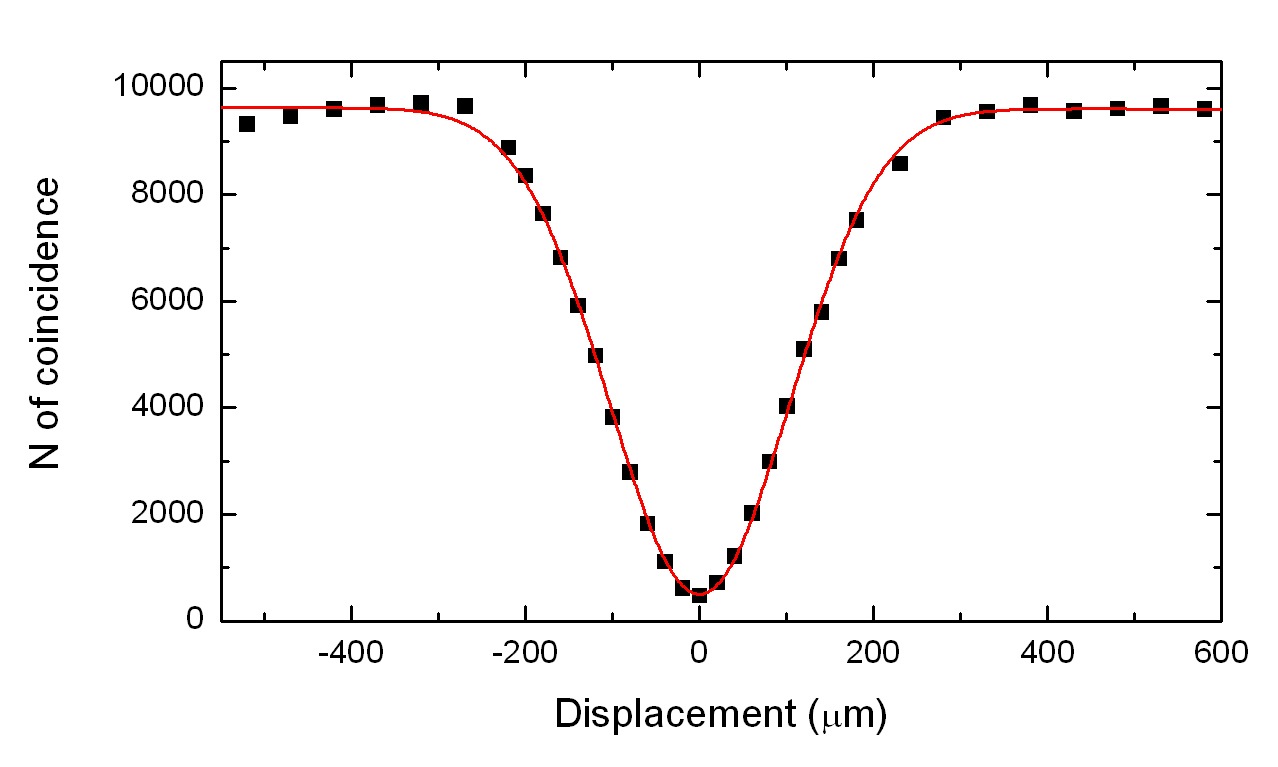}
\caption{Quantum interference in an integrated waveguide coupler. Error bars are smaller than the data points.} 
\label{fig2}
\end{center}
\vspace{-0.7cm}
\end{figure}

Figure 2 shows the classic signature of quantum interference: a dramatic dip in the rate of detecting two photons at each output of a directional coupler near zero delay in relative photon arrival time \cite{ho-prl-59-2044}. The raw visibility \footnote{$V\equiv (max-min)/max$}  $V=94.8\pm0.5\%$ is a measure of the quality of the interference and demonstrates very good quantum behavior of photons in an integrated optics architecture. 

Figure 3A shows the measured non-classical visibility for 10 couplers on a single chip with a range of design $\eta$'s. The observed behavior is well explained by the theoretical curves which include a small amount of residual mode mismatch and an offset of $\delta\eta=3.4\pm0.7\%$ from the design ratio. Similar behavior was observed for a second set of devices on a second chip (not shown). The waveguides are designed such that the cut-off wavelength for higher order modes is very near to the design wavelength in order to maintain a large waveguide core size. This suggests that the small residual mode mismatch could be in the spatial mode overlap since weakly guided higher order modes may propagate across the relatively short devices. However it is inherently difficult to identify which degree of freedom in which mode mismatch occurs \cite{ro-pra-72-032306}. These results demonstrate the high yield and excellent reproducibility of the devices. 

\begin{figure}[t]
\begin{center}

\includegraphics[width=0.4\textwidth]{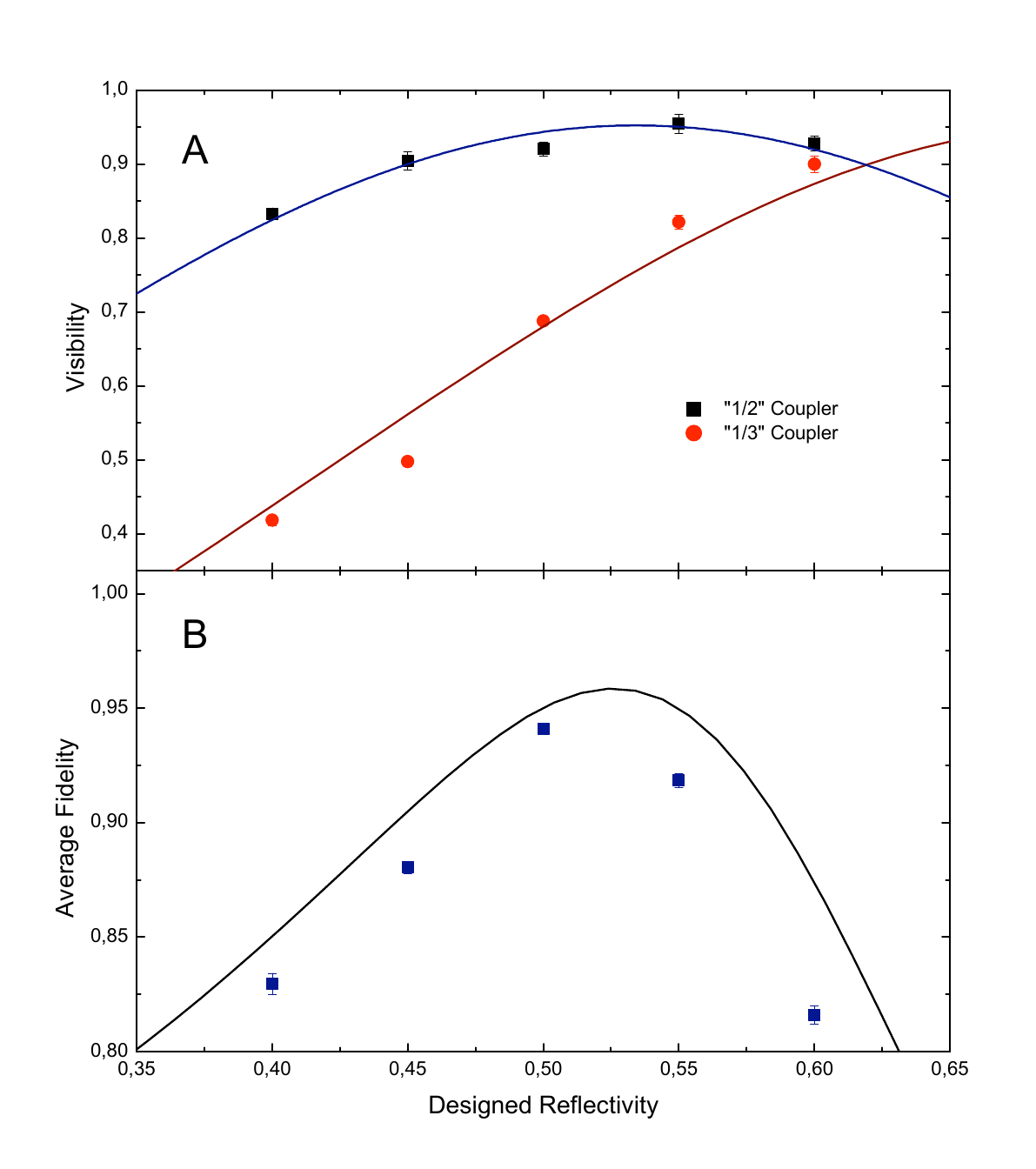}
\vspace{-0.3cm}
\caption{Two photon quantum interference on chip. ({\bf A}) Quantum interference visibility at
``1/2" and ``1/3"couplers  that compose a CNOT gate (where the ``1/2" couplers range from $\eta=0.4-0.6$ and the ``1/3" couplers are 2/3 this value: $\eta(``1/3")=0.27-0.4$). The fit to the ``1/2" data includes an offset in the coupling ratio $\delta\eta$ and mode mismatch $\varepsilon$ as free parameters. The same values are used for the ``1/3" theoretical curve. ({\bf B}) Logical basis fidelity for each of the CNOT gates. The solid curve corresponds to a model including only these values of $\varepsilon$ and $\delta\eta$. The model does not include the effect of classical interference, which explains the offset.} \label{fig3}
\end{center}
\vspace{-0.7cm}
\end{figure}

General photonic quantum circuits require both quantum and classical interference, and their combination for conditional phase shifts \cite{kn-nat-409-46}. An ideal device for testing all of these requirements is the entangling controlled-NOT (CNOT) logic gate shown in Fig. 1C \cite{ra-pra-65-062324,ho-pra-66-024308}, which has previously been experimentally demonstrated using bulk optics \cite{ob-nat-426-264,ob-prl-93-080502,la-prl-95-210504,ki-prl-95-210505,ok-prl-95-210506}. The control $C$ and target $T$ qubits are each encoded by a photon in two waveguides and the success of the gate is heralded by detection of a photon in both the control and target outputs, which happens with probability 1/9. 
We note that the waveguide implementation of this gate is essentially a direct writing onto the chip of the theoretical scheme presented in \cite{ra-pra-65-062324}.

To allow for possible design and fabrication imperfections we designed and fabricated on the same chip several CNOT devices with``1/2" couplers ranging from $\eta=0.4-0.6$, and correspondingly ``1/3" couplers ranging from $\eta=0.27-0.4$ (\emph{i.e.} 2/3 of the ``1/2" couplers). The  quantum interference measurements described above (Fig. 3B) show that the devices are in fact very close to the design $\eta$: $\delta\eta=3.4\pm0.7\%$. Note that to measure the ``1/2" couplers we sent single photons into the $T_0$ and $T_1$ inputs, and collected photons from the $C_1$ and $V_B$ outputs (and the reverse for the other ``1/2" coupler); the ``1/3" data are for the couplers between the $C_0$ and $V_A$ waveguides (see Fig. 1C). 

\begin{figure*}[t]
\begin{center}
\vspace{-0.8cm}
\center{\includegraphics[width=0.75\textwidth]{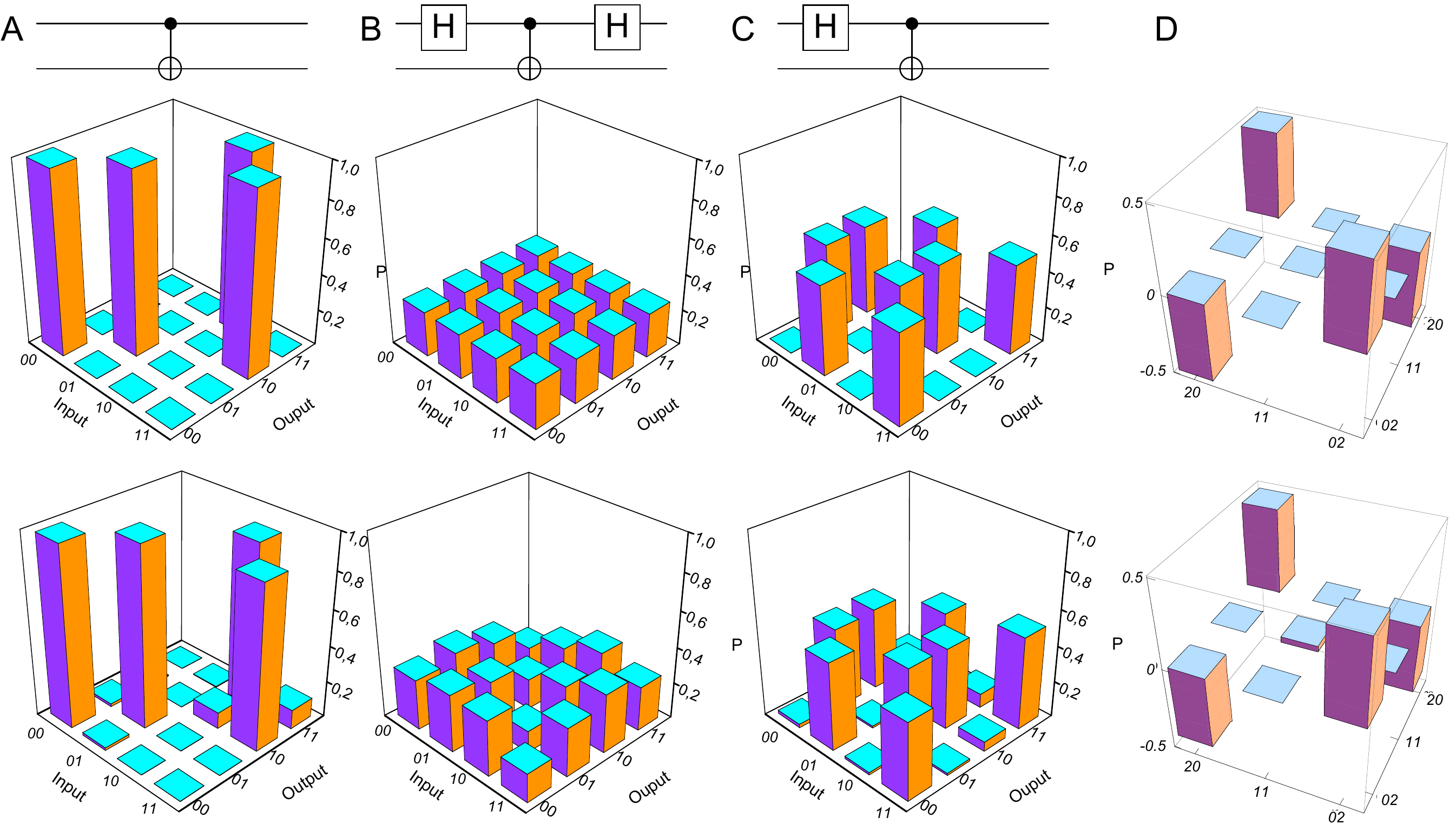}}
\vspace{-0.4cm}
\caption{Characterization of integrated quantum photonic circuits.  Ideal and
measured truthtables for a CNOT circuit ({\bf A}); a CNOT with two additional H gates ({\bf B}); and a CNOT with one additional H gate. {\bf D} The ideal and estimated density matrix for the maximally path entangled state $(|20\rangle-|02\rangle)/\sqrt{2}$ } \label{fig4}
\end{center}
\vspace{-0.4cm}
\end{figure*}

Focusing on the CNOT device with nominally 50:50 couplers, we input the four computational basis states $|0\rangle_C|0\rangle_T$, $|0\rangle_C|1\rangle_T$, $|1\rangle_C|0\rangle_T$, and $|1\rangle_C|1\rangle_T$ and measured the probability of detecting each of the computational basis states at the output (Fig. 4A). The excellent agreement for the $|0\rangle_C$ inputs (peak values of $98.5\%$) is a measure of the classical interference in the target interferometer and clearly demonstrates that the waveguides are stable on a subwavelength scale---a key advantage arising from the monolithic nature of an integrated optics architecture. The average logical basis fidelity $F=94.3\pm0.2\%$ is the highest yet reported for any entangling logic gate, not just for photons, but in any experimental architecture. The fidelities for the other four devices (with different $\eta$'s) are lower (Fig. 3B), as expected.

To directly confirm coherent quantum operation and entanglement in our devices we launched pairs of photons into the $T_0$ and $T_1$ waveguides. This state should ideally be transformed at the first 50:50 coupler as follows: 
\begin{equation}
|11\rangle_{T_0T_1}\rightarrow(|20\rangle_{T_0T_1}-|02\rangle_{T_0T_1})/\sqrt{2},
\label{hom}
\end{equation}
\emph{i.e.} a maximally path entangled superposition of two photons in the top waveguide and two photons in the bottom waveguide. A very low rate of detecting a pair of photons at the $C_1$ and $V_A$ outputs, combined with a high rate of detecting two photons in either of these outputs (via a pair of cascaded SPCMs) confirmed that the state was predominantly composed of $|20\rangle$ and $|02\rangle$ components, but did not indicate a coherent superposition. At the second 50:50 coupler between the $T_0$ and $T_1$ waveguides the reverse transformation of Eq. \ref{hom} should occur, provided the minus superposition exists. A high rate of detecting photon pairs at the $T_0$ and $T_1$ outputs combined with a low rate of detecting two photons in either of these outputs confirmed this transformation. 
From each of these measured count rates we were able to estimate the two-photon density matrix (Fig. 4D). The fidelity with the maximally path entangled state $|20\rangle-|02\rangle$ is $>92\%$ \footnote{Note that we cannot measure the four ``zero" coherences in the density matrix (although they are limited by the small $|11\rangle\langle11|$ population), nor distinguish between non-maximal coherences and rotated coherences with imaginary components for the $|20\rangle\langle02|$ and $|02\rangle\langle20|$ terms. However, neither of these effects change the state fidelity. We have assumed a worst case scenario throughout---\emph{i.e.} we assume that the $|11\rangle$ component inside the interferometer makes no contribution to two photon detection in $T_0$ or $T_1$.}. This high fidelity generation of the lowest order maximally path entangled state combined with confirmation of the phase stability of the superposition demonstrates the applicability of integrated devices for quantum metrology applications.

Finally, we tested the simple quantum circuits shown in Figs. 4B and C, consisting of a CNOT gate and Hadarmard H gates---$|0\rangle\rightarrow|0\rangle+|1\rangle$; $|1\rangle\rightarrow|0\rangle-|1\rangle$---each implemented with a 50:50 coupler between the $C_0$ and $C_1$ waveguides. In both cases we observe good agreement with the ideal operation, as quantified by the average classical fidelity between probability distributions \cite{pr-prl-92-190402,ra-pra-73-012113}: $97.9\pm 0.4\%$ and $91.5\pm 0.2\%$, respectively. The device shown in Fig. 4B should produce equal superpositions of the four computation basis states $|00\rangle\pm|01\rangle\pm|10\rangle\pm|11\rangle$ and that shown in Fig. 4C should produce the four maximally entangled Bell states $\Psi^{\pm}\equiv|01\rangle\pm|10\rangle$ and $\Phi^{\pm}\equiv|00\rangle\pm|11\rangle$. While this cannot be confirmed directly on-chip, the above demonstrations of excellent logical basis operation of the CNOT and coherent quantum operation give us great confidence.

We note that previous bulk optical implementations of these photonic quantum circuits circuit, as well as other circuits, have required the design and implementation of sophisticated interferometers. Constructing such interferometers have been a major obstacle to the realization of photonic quantum circuits. The results presented here show that this problem can be drastically reduced by using waveguide devices: it becomes possible to directly write the theoretical ``blackboard sketch" onto the chip, without requiring sophisticated interferometers.

We have demonstrated high fidelity integrated implementations of each of the key components of photonic quantum circuits, as well as several small-scale circuits. This opens the way for miniaturizing, scaling, and improving the performance of photonic quantum circuits for both future quantum technologies and the next generation of fundamental quantum optics studies in the laboratory. 
\footnote{We thank A. Clarke, J. Fulconis, A. Gilchrist, A. Laing, S. Lardenois, G. Maxwell, and A. Stefanov for helpful discussions. This work was supported by the US Disruptive Technologies Office (DTO), the UK Engineering and Physical Sciences Research Council (EPSRC), the UK Quantum Information Processing Interdisciplinary Collaboration (QIP IRC),  and the Leverhulme Trust.}

\end{document}